# 140 Gbaud On-Off Keying Links in C-Band for Short-Reach Optical Interconnects

Oskars Ozolins[1], José Manuel Estaran[2], Aleksejs Udalcovs[1], Filipe Jorge[3], Haik Mardoyan[2], Agnieszka Konczykowska[3], Muriel Riet[3], Bernadette Duval[3], Virginie Nodjiadjim[3], Jean-Yves Dupuy[3], Xiaodan Pang[4],[1], Urban Westergren[4], Jiajia Chen[4], Sergei Popov[4], Sébastien Bigo[2]

[1] Networking and Transmission Laboratory, RISE Acreo AB, Sweden, oskars.ozolins@ri.se
[2] Nokia Bell Labs, 7 route de Villejust, 91620 Nozay, France
[3] III-V Lab, joint lab of Nokia Bell Labs, Thales Research and Technology and CEA Leti, 1 av. Augustin Fresnel, 91676 Palaiseau, France
[4] School of EECS and SCI, KTH Royal Institute of Technology, Kista, Sweden

**Abstract** *We demonstrate 140 Gbaud intensity modulated direct detection dispersion-uncompensated links with Mach Zehnder modulator and distributed feedback travelling-wave electro-absorption modulator over 5500 and 960 meters of standard single mode fibre, respectively, enabled by compact packaged ultra-high speed InP-based 2:1-Selector.*

## Introduction

Datacentres experience massive traffic growth due to the explosive amount of information to be stored, transmitted and processed[1,2]. Short-reach optical interconnects operating at 100 Gbaud and beyond per lane are required to confront the growing bandwidth requirements in datacentres[3]. Scaling the baudrate even higher to reduce the cost per bit in highly parallel systems is the way to go[4]. By extensive integration of opto-electronic components one achieves a reduced size and a good power efficiency[3,5]. Furthermore, it is important to reduce the complexity by employing intensity modulation and direct detection (IM/DD). Several attractive technological solutions for the modulator are proposed[5,6]. The C-band enables proportional efficiency in bidirectional bandwidth for intra-datacentres thanks to mature and efficient wavelength division multiplexing technology. On-off keying (OOK) is the simplest way to modulate information, allowing for ultra-high speed and power- and cost-effective electronics[7]. This path is possible thanks to indium phosphide (InP) ultra-high-speed components with bandwidth beyond 70 GHz[8-11] enabling beyond 100 Gbaud per lane transmitters[7] for inter-datacentre links.

In this paper we demonstrate 140 Gbaud on-off keying intra-datacentre links in C-band using Mach Zehnder modulator (MZM) and distributed feedback travelling wave electro absorption modulator (DFB-TWEAM). In case of back to back (b2b) for DFB-TWEAM we achieve performance below 7 % overhead (OH) hard decision forward error correction (HD-FEC) code bit error rate (BER) threshold[12] without adaptive equalization. High speed electrical signal generation is realized by a compact packaged high-speed InP-based 2:1-Selector. In addition, we extend our previous demonstration[7] by transmitting with MZM over 5500 meters of standard single mode fibre (SSMF) without amplification and dispersion compensation.

## Experimental setup

Figure 1 shows the experimental setup for demonstration of 140 Gbaud on-off keying optical links. It consits of electrical signal generation, different modulators, optical link and receiver. To generate high data rate signals we use the electrical time division multiplexed (ETDM) 2:1-Selector (SEL)[9]. The SEL is fabricated in 0.7 µm InP double heterojunction bipolar transistor (DHBT) technology over 1.2x1.5 mm². The power consumption is 0.5 to 0.8 W, for a differential output amplitude of 250 to 730 mV. Seven power and control supplies allow fine optimization of the output signals' amplitude and jitter, so that we can optimize different bias setting for MZM and DFB-TWEAM. Eye diagram measurements (with a 70-GHz oscilloscope) at 140 Gbaud are presented in Fig. 2(b). The low-pass filtering effects are mainly due to the limitations of the recording equipment as well as packaging.

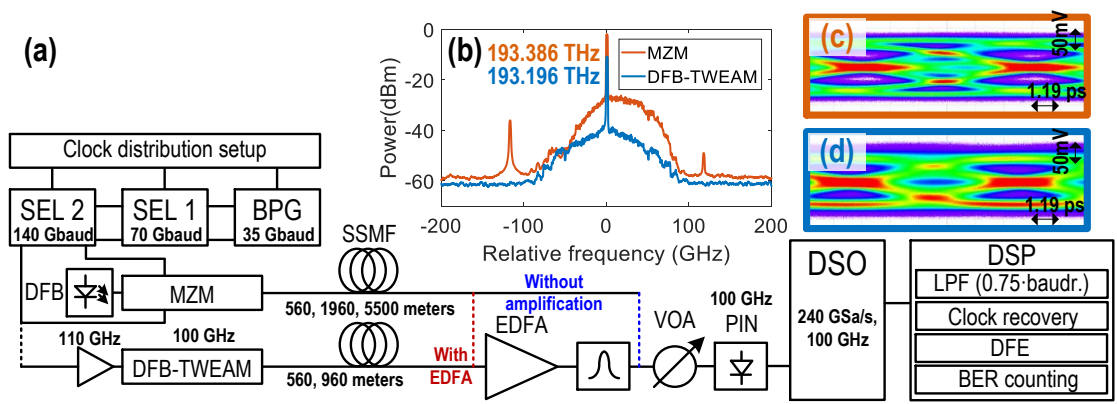

**Fig. 1:** (a) Experimental setup for the 140 Gbaud IM/DD links. (b) 140 Gbaud OOK optical spectra after MZM and DFB-TWEAM output; 140 GBaud OOK eye diagram for (c) MZM and (d) DFB-TWEAM in back to back at DSO before offline processing.

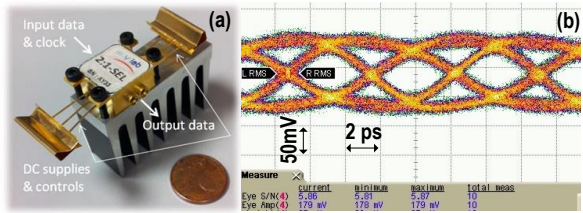

**Fig. 2:** (a) InP DHBT 2:1-Selector package photograph; (b) output OOK eye diagrams at 140 Gbaud.

We are using two types of modulators: distributed feedback travelling wave electro absorption modulator and Mach Zehnder modulator. The DFB-TWEAM has 3 dB bandwidth beyond 100 GHz with less than 2 dB ripple in the pass band which indicates high phase linearity[10]. We use 115 mA current for DFB and -1.85 V bias voltage for TWEAM. To achieve sufficient single ended driving signal amplitude, we use an IAF InP DHBT-based modulator driver module with a gain of 16 dB and 3 dB bandwidth of 110 GHz[11]. The main performance limitations come from packaging and clock distribution at the transmitter side. In case of MZM (FTM7935EZ, 43 Gbit/s) we are using differential driving signal without external driver amplifier. Furthermore, we use distributed feedback (DFB) laser with 13 dBm output power. We investigate two link configurations without and with erbium doped fibre amplifier (EDFA) with various SSMF lengths (see Fig. 1(a)). We use EDFA as preamplifier instead of a transimpedance amplifier to reduce bandwidth limitation for high-speed signals.

We use a high-speed bit pattern generator (BPG) from SHF to generate two 35 Gbaud electrical signals with $2^{15}-1$ long pseudo-random bit sequences (PRBS). Then we feed a pair of 2:1 SEL in series. The first 2:1-Selector (SEL1 in Fig. 1(a)) yields output signal at 70 Gbaud and second 2:1-SEL (SEL2) at 140 Gbaud. We are observing bandwidth limitations due to its packaging. The clocks required to drive the components were produced from a master clock, distributed after multiplication/division up to baudrate/8 frequency for BPG, baudrate/4 for SEL1 and baudrate/2 for SEL2. The modulated optical spectra are displayed in Fig. 1(b). Receiver consists of variable optical attenuator (VOA) and commercial 100 GHz bandwidth photodiode. The received signal is then sampled by a Teledyne LeCroy Digital Storage Oscilloscope (DSO) with 100 GHz of bandwidth operating at 240 GSa/s and stored for offline processing. The receiver digital signal processing (DSP) routine consists of the standard algorithm stack: clock recovery, resampling, equalization, and error counting. The signal is equalized with symbol-spaced decision-feedback equalizer (DFE) using different configuration of feed-forward taps (FFT) and feedback taps (FBT) to overcome inter-symbol interference in presence of noise (see Fig. 1(a)).

**Results and discussions**

FEC code with 7 % OH is considered (pre-FEC BER limit at $5\cdot10^{-3}$)[12] for result analysis. We evaluate various transmission distances and equalizer taps for each of the link configurations.

In Fig.3 we are showing BER performance for link configuration with MZM without amplification and dispersion compensation. BER curves are obtained using 6-FFT&6-FBT and 12-FFT&6-FBT DFE. No amplification was used in this case. We manage to achieve performance below 7 % HD-FEC up to 5500 meters SSMF which was limited by link power budget. We attribute achieved reach performance to modulator chirp and dispersion interaction. Furthermore, one could observe clearly asymmetrical optical spectra on Fig.1(b) which shows partly single sideband modulation.

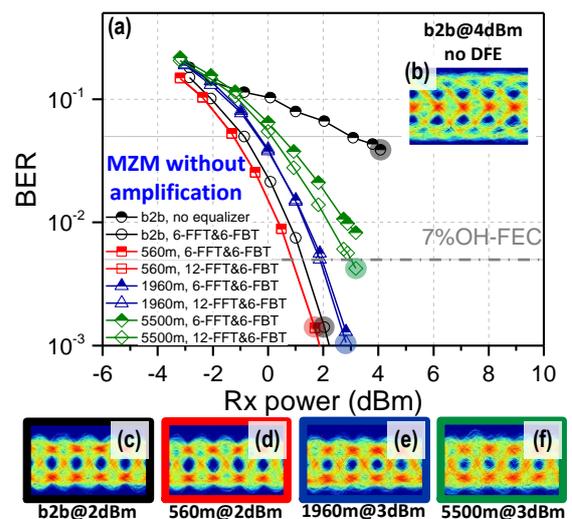

**Fig. 3:** Performances of (a) 140-GBaud OOK link using MZM without EDFA for different equalizers; (b) eye diagram without equalizer; (c)-(f) eye diagrams with DFE for different links shown as insets.

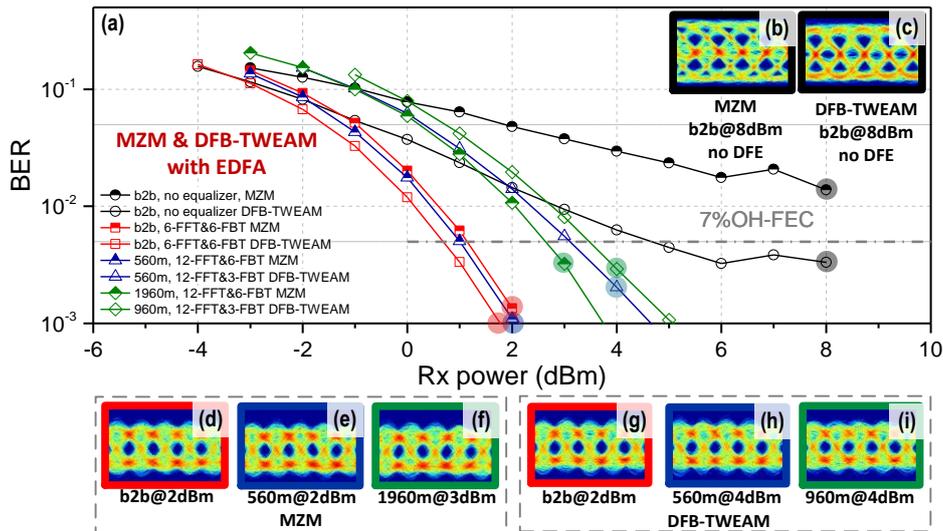

**Fig. 4:** Performances of (a) 140-GBaud OOK link using MZM and DFB-TWEAM with EDFA for several number of equalizers; (b) eye diagram without equalizer; (c)-(i) eye diagrams with DFE for different links shown as insets.

We extend the configuration with EDFA as preamplifier instead of a transimpedance amplifier to reduce bandwidth limitation for high-speed signals and compensate for -1 dBm output power for DFB-TWEAM with modulation. We present results in Fig.4. We manage to achieve performance below 7% OH HD-FEC[12] BER threshold without adaptive equalization thanks to the extraordinary good signal-generation quality that is delivered by integrated InP DHBT electronics together with DFB-TWEAMs modules. As well in this case MZM shows longer reach thanks to chirp and dispersion interaction combined with partly single sideband modulation.

### Conclusions
We demonstrate IM/DD C-band links with DFB-TWEAM and MZM without dispersion compensation. We show 5500 meters reach of SSMF with MZM over without amplification and dispersion compensation.

### Acknowledgements
We thank Teledyne LeCroy and SHF for the loan of equipment. The work was supported by the European Commission through H2020 projects QAMeleon (no. 780354) and NEWMAN (no. 752826), the Swedish Research Council project PHASE (no. 2016-04510), the Swedish ICT–TNG project SCENE, the VINNOVA projects SENDATE-FICUS and SENDATE-EXTEND.